\documentclass[reprint,
superscriptaddress,
amsmath,amssymb,
aps,nofootinbib,
prd,onecolumn
]{revtex4-2}

\usepackage{graphicx}
\usepackage[usenames,dvipsnames]{xcolor}
\definecolor{bluecite}{HTML}{0875b7}
\usepackage[colorlinks=true,urlcolor=Emerald,linkcolor=bluecite,citecolor=bluecite]{hyperref}
\usepackage{acronym}
\usepackage{MnSymbol}
\usepackage{physics}
\usepackage{float}

\def \M{\mathcal{M}}

\begin{document}
	
	\title{A numerical analysis of Araki-Uhlmann relative entropy in Quantum Field Theory}
	
	\author{Marcelo S.  Guimaraes,}
	\email{msguimaraes@uerj.br}
	\affiliation{UERJ - Universidade do Estado do Rio de Janeiro,	Instituto de Física - Departamento de Física Teórica - Rua São Francisco Xavier 524, 20550-013, Maracanã, Rio de Janeiro, Brazil}
	
	\author{Itzhak Roditi}
	\email{roditi@cbpf.br}
	\affiliation{CBPF - Centro Brasileiro de Pesquisas Físicas, Rua Dr. Xavier Sigaud 150, 22290-180, Rio de Janeiro, Brazil}
	
	\author{Silvio P. Sorella}
	\email{silvio.sorella@fis.uerj.br}
	\affiliation{UERJ - Universidade do Estado do Rio de Janeiro,	Instituto de Física - Departamento de Física Teórica - Rua São Francisco Xavier 524, 20550-013, Maracanã, Rio de Janeiro, Brazil}
	
	\author{Arthur F. Vieira}
	\email[]{arthurfvieira@if.ufrj.br}
	\affiliation{UERJ - Universidade do Estado do Rio de Janeiro,	Instituto de Física - Departamento de Física Teórica - Rua São Francisco Xavier 524, 20550-013, Maracanã, Rio de Janeiro, Brazil}
	\affiliation{Instituto de Física, Universidade Federal do Rio de Janeiro,	21.941-909, Rio de Janeiro, RJ, Brazil}
	
	\date{\today}
	
	\begin{abstract}
		We numerically investigate the Araki-Uhlmann relative entropy in Quantum Field Theory, focusing on a free massive scalar field in $1+1$-dimensional Minkowski spacetime. Using Tomita-Takesaki modular theory, we analyze the relative entropy between a coherent state and the vacuum state, with several types of test functions localized in the right Rindler wedge.  Our results confirm that relative entropy decreases with increasing mass and grows with the size of the spacetime region, aligning with theoretical expectations.
	\end{abstract}
	\maketitle
	
	\tableofcontents
	
	\section{Introduction}\label{sec:intro}

	The evaluation of entanglement entropy in relativistic Quantum Field Theory (QFT) presents a profound challenge, engaging a variety of  technically demanding methods \cite{Nishioka:2018khk}. Central to the discussion of quantum information measures in Quantum Mechanics (QM)  and QFT is the von Neumann entropy, which quantifies the uncertainty or information content of a quantum state. For a given density matrix $\rho$ that describes the state of the system, the von Neumann entropy is defined as 
	\begin{equation}
		S_{\rm vN}(\rho) = -\Tr(\rho \log \rho),
	\end{equation}
	which generalizes the classical statistical definition in terms of probability distributions. This concept finds a natural implementation in the study of pure bipartite systems, where the entanglement entropy $S_{\rm E}$ of a subsystem is derived from the von Neumann entropy of its reduced density matrix. This reduced density matrix is obtained by tracing out the degrees of freedom of the complementary subsystem. If the entire system is described by a product state, the reduced density matrices will be pure, resulting in a vanishing von Neumann entropy. Conversely, if the subsystems are entangled, the von Neumann entropy will be non-zero, making it an useful measure of entanglement.
	
	In QFT, the entanglement entropy emerges as a pivotal measure, reflecting the degree of quantum entanglement between different spatial regions, see \cite{Holzhey:1994we,Calabrese:2004eu,Calabrese:2009qy} for early works and \cite{Casini:2022rlv,Berges:2017hne,Schrofl:2023hnz,Katsinis:2023hqn} for recent accounts. However, it is inherently plagued by ultraviolet (UV) divergences due to the infinite number of degrees of freedom near the boundary of the regions. In a QFT  in $(d+1)-$dimensions with a UV cutoff $\epsilon$, the entanglement entropy of a region $V$ typically displays a structure composed of power-law and logarithmic divergent terms of the form \cite{Casini:2022rlv,Marolf:2016dob}
	\begin{equation}
		S_{\rm E}(V) = S_0(V)+ g_{d-1}[\partial V]\epsilon^{-(d-1)} + \cdots + g_1[\partial V]\epsilon^{-1}+g_{0}[\partial V]\ln \epsilon + \mathcal{O}(\epsilon).
	\end{equation}
	Here, $S_0(V)$ represents the finite part of the entanglement entropy, $g_{i}$ are local functionals of the boundary $\partial V$ and may depend on the specifics of the field theory. They are proportional to the $(i - 1)$-th power of a characteristic length scale of $V$. The term with the highest power of $\epsilon$ is associated with $g_{d-1}$, which has the dimension of $[\text{length}]^{d-1}$, corresponding to an area law. The terms containing $g_i$ for $i > 0$ are not universal, as they depend on the chosen regularization scheme. However, the logarithmic divergence coefficient, $g_0$, is considered universal and independent of the cutoff. These divergences arise from the short-distance correlations near the entangling surface once we trace out the degrees of freedom outside the region $V$ and are generally state-independent (however, see \cite{Marolf:2016dob}). See also \cite{Pagani:2018mke}, which discusses the UV divergences of the entanglement entropy in the context of the asymptotically safe quantum theory of gravity.
	
	A related concept is the relative entropy. In QM, the relative entropy between two density matrices $\rho$ and $\sigma$ is defined as
	\begin{equation}\label{eq.rel.entropy}
		S(\rho | \sigma) = \Tr(\rho \log \rho - \rho \log \sigma),
	\end{equation}
	which is non-negative and zero if and only if $\rho = \sigma$. The relative entropy measures the distinguishability between two states. Operationally, if we have a state $\sigma$ and make $N$ measurements to check how closely the outcomes align with the expectations of another state $\rho$ (interpreted as a theoretical model), the probability $p$ of obtaining similar results decays exponentially with $N$ if the states differ, following $p \sim \exp(-N S(\rho | \sigma))$ \cite{Hiai:1991mxv,Casini:2022rlv}. Recent studies of relative entropy in QFT  can be found in \cite{Hollands:2019czd,DAngelo:2021yat,Ciolli:2021otw,Galanda:2023vjk,Garbarz:2022wxn}, while Refs. \cite{Floerchinger:2020ogh,Dowling:2020nxc} explore the derivation of thermodynamics and \cite{Jafferis:2014lza,Jafferis:2015del,Verlinde:2019ade,Bousso:2020yxi} discuss the role of relative entropy in the context of the gauge-gravity duality. Note that, being essentially a difference between entropies, the UV divergence of entropies in QFT cancels in the relative entropy expression.
	
	Using the Tomita-Takesaki modular theory \cite{tomita1967canonical,Takesaki:1970aki}, one can construct the Araki-Uhlmann relative entropy \cite{Araki:1975zw,Araki:1976zv,Uhlmann:1976me}, a particularly useful generalization of the relative entropy to von Neumann algebras that uses powerful tools such as the Haag-Kastler formulation of QFT and the Bisognano-Wichmann theorems for wedge regions, see \cite{Witten:2018zxz} for a general account. It is defined between two states $\ket{\Psi}$ and $\ket{\Phi}$ associated with a von Neumann algebra $\M$ and is given by
	\begin{equation}
		S(\Psi | \Phi) = -\bra{\Psi}  \log \Delta_{\Psi|\Phi} \ket{\Psi},
	\end{equation}
	where $\Delta_{\Psi|\Phi}$ is the relative modular operator. The Araki-Uhlmann relative entropy serves as a finite, state-independent measure, and it has been applied effectively in understanding the properties of entanglement in various QFT setups \cite{Casini:2017roe,Hollands:2019ajl,Casini:2019qst,Longo:2019mhx,Dowling:2020nxc,DAngelo:2021yat,Morinelli:2021nsx,Casini:2022bsu,DAngelo:2023trx,Schrofl:2023hnz,Frob:2023tcj,Katsinis:2024gef,Abate:2024xyb,Longo:2024fkb}. Interestingly, in QM, depending on the type of von Neumann algebra, the Araki-Uhlmann formula reduces to Eq.~\eqref{eq.rel.entropy} in terms of density matrices, see, for instance, \cite{Frob:2023tcj} and Sec. IV of \cite{Witten:2018zxz}.
	
	In this work, we investigate the Araki–Uhlmann relative entropy between a coherent state and the vacuum state of a massive scalar field in (1+1)-dimensional Minkowski spacetime. Despite the conceptual significance of the relative entropy, explicit evaluations in field-theoretic settings often demand careful numerical treatment. To this end, by using modular theory and the Tomita–Takesaki framework, we develop a numerical approach to explore the positivity of this entropy and its behavior with respect to key parameters, such as mass and the size of spacetime regions.
	
	A crucial aspect of our analysis is the introduction of test functions, which is essential for obtaining a well-defined expression for relative entropy in QFT. This requirement arises naturally from the distributional nature of quantum fields. Despite this necessity, fundamental properties of relative entropy — such as positivity and its monotonic increase with the size of the spacetime region — are expected to remain independent of the specific choice of test function.
	
	To explicitly address this issue, we conduct a detailed numerical analysis by employing four different types of test functions. Our results confirm the universality of the expected entropy properties, reinforcing the robustness and versatility of relative entropy in characterizing entanglement.
	
	Additionally, we examine the dependence of entropy on the mass parameter. We demonstrate that, across all cases considered, the entropy decreases as the mass increases. To the best of our knowledge, this finding is both novel and significant, providing new insights into the behavior of entropy in QFT.
	
The principal findings can be summarized as follows:

\begin{itemize}
	\item We have shown in an explicit way that the behavior of the relative entropy does not depend on the specific choice of the test functions. This is done by showing, through a numerical 
	approach, that the behavior of the entropy remains unchanged upon using four different types of test functions. 
	
	\item Relative entropy decreases monotonically with increasing mass. While this trend aligns with theoretical expectations,  we are not aware of any existing formal proof of this result. Also here, the same decreasing behavior with respect to the  mass parameter  is common to all employed  test functions. Our results thus provide the first explicit confirmation across different test functions.
	
	\item Finally, we provide an explicit check of the monotonic increase of the entropy with the size of the spacetime regions under considerations. 
\end{itemize}

These findings reinforce our current understanding of the Araki-Uhlmann entropy and offer concrete numerical evidence supporting its robustness in free field theories.
	
	The paper is structured as follows: Sec.~\ref{sec.background} reviews the basic concepts of Araki-Uhlmann relative entropy. Sec.~\ref{sec.ftest} details the construction of the test function. Sec.~\ref{ns} focuses on the numerical setup and presents the results. Sec.~\ref{conc} concludes the paper. Appendix \ref{appA} provides a brief summary of some key notions of canonical quantization of the massive scalar field needed for readability.

	\section{Generalities on the Araki-Uhlmann relative entropy}\label{sec.background}

	In this section, we review the main properties of the Araki-Uhlmann relative entropy \cite{Araki:1976zv,Witten:2018zxz} in QFT. The first step is to consider a von Neumann algebra ${\cal M}$ equipped with two cyclic and separating states: $\ket{\Psi}, \ket{\Omega}$. Araki-Uhlmann relative entropy is thus defined by 
	\begin{equation} 
		S(\Psi |\Omega) = - \bra{\Psi}\log \Delta_{\Psi| \Omega}  \ket{\Psi} .\label{ent}
	\end{equation}
	As stated in the Introduction, here $ \Delta_{\psi| \Omega}$ is the relative Tomita-Takesaki modular operator \cite{tomita1967canonical,Takesaki:1970aki,Witten:2018zxz} and is obtained by means of the relative anti-linear operator $s_{\Psi| \Omega}$, whose action is defined by the closure of the map 
	\begin{equation} 
		s_{\Psi |\Omega} \; a \ket{\Psi} = a^{\dagger} \; \ket{\Omega} \;, \qquad \forall a \in {\cal M} \;. \label{Stt}
	\end{equation} 
	The polar decomposition of $ s_{\Psi| \Omega} $  gives thus  $\Delta_{\Psi| \Omega}$, namely 
	\begin{equation} 
		s_{\Psi| \Omega} = J_{\Psi |\Omega} \;  \Delta_{\Psi| \Omega}^{1/2} \;, \label{pd}
	\end{equation}
	with $J_{\Psi|\Omega}$ being the anti-unitary relative modular conjugation. The operator $ \Delta_{\Psi |\Omega}$ is self-adjoint and positive definite. 
	
	Following \cite{Ciolli:2019mjo}, an operational way of facing expression \eqref{ent} can be achieved through the spectral decomposition of $ \Delta_{\Psi| \Omega}$, yielding 
	\begin{equation} 
		S(\Psi |\Omega) = i \frac{d}{ds} \bra{\Psi}\;  \Delta_{\Psi| \Omega}^{is} \; \ket{\Psi} \Big|_{s=0} \;, \label{ds}
	\end{equation}
	where the unitary operator 
	\begin{equation} 
		\Delta_{\Psi |\Omega}^{is} = e^{i s \log \Delta_{\Psi| \Omega} } \;, \qquad s \in {\mathbb R}  \;, \label{mf}
	\end{equation}
	is known as the modular flow \cite{Takesaki:1970aki,Witten:2018zxz}. 
	
	To proceed, one has to specify the states $(\ket{\Psi}, \ket{\Omega})$. As already mentioned, the state $\ket{\Psi}$ is taken to be a coherent state localized in the right wedge ${\cal W}_{R} = \left\{ \;\vb*{x}=(t,x)\;, x\ge |t| \;  \right\} $, while $|\Omega\rangle$ denotes the vacuum state. Coherent states are obtained by acting with the Weyl operators on the vacuum state, namely 
	\begin{equation} 
		\ket{\Psi}= {\cal A}_f \ket{\Omega} = e^{i \varphi(f)} |\Omega\rangle  \;, \label{coh}
	\end{equation}
	where $\varphi(f)$ is the smeared scalar field (see Appendix \eqref{appA}) and $f(\vb*{x})$ is a smooth test function whose compact support is localized in ${\cal W}_R$. The explicit form of $f(\vb*{x})$ will be discussed in detail in Sec.~\ref{sec.ftest}.  
	The Weyl operators are unitary operators fulfilling the following properties: 
	\begin{eqnarray} 
		{\cal A}_f  {\cal A}_g & = & e^{-\frac{i}{2} \Delta_{PJ}(f,g)} \; {\cal A}_{(f+g)} \;, \nonumber \\
		{{\cal A}}_f ^{\dagger} { {\cal A}} _f & = & 1 \;, \qquad {\cal A}_f {\cal A}_f ^\dagger =1 \;, \qquad {\cal A}_f^{\dagger} = {\cal A}_{-f} \;, \label{Wa}
	\end{eqnarray} 
	and 
	\begin{equation} 
		\langle \Omega | \; {\cal A}_f \; |\Omega \rangle = e^{-\frac{1}{2} ||f||^2 } \;, \label{vev}
	\end{equation}
	where $||f||^2$ stands for the norm induced by the Lorentz invariant inner product 
	\begin{equation} 
		\langle f | g \rangle = \int \frac{dk}{2\pi} \frac{1}{2\omega_k} f^{*}(\omega_k,k) g(\omega_k,k) = \frac{i}{2}\Delta_{PJ}(f,g) + H(f,g)  \;. \label{ip}
	\end{equation}
	Here, $f(\omega_k,k)$ is the Fourier transform of $f(t,x)$ and $\Delta_{PJ}(f,g)$, $H(f,g)$ are the smeared Pauli-Jordan and Hadamard distributions, given in Eqs.~\eqref{mint} and \eqref{PJH}. 
	
	A useful aspect of examining the Araki-Uhlmann relative entropy between a coherent state and the vacuum state lies in the notable relation  \cite{Casini:2019qst,Frob:2024ijk}
	\begin{equation} 
		\Delta_{\Psi| \Omega}^{is} = \Delta_{\Omega}^{is} \;, \label{mde}
	\end{equation} 
	where $\Delta_{\Omega}^{is}$ is the  flow of the Tomita-Takesaki modular operator for the vacuum state $\ket{\Omega}$ \cite{Takesaki:1970aki,Witten:2018zxz} such that
	\begin{eqnarray} 
		s_{\Omega} \; a |\Omega\rangle &  =  & a^{\dagger} |\Omega\rangle \;, \qquad \forall a \in {\cal M}  \;, \nonumber \\
		s_{\Omega} & = & J_{\Omega} \Delta_{\Omega}^{1/2}  \;. \label{som}
	\end{eqnarray}
	In particular, from the Bisognano-Wichmann results \cite{Bisognano:1975ih}, the action of the modular flow $\Delta_{\Omega}^{is}$ on the Weyl operators is known, being given by 
	\begin{equation} 
		\Delta_{\Omega}^{is}\; {\cal A}_f \; \Delta_{\Omega}^{-is} = e^{i \varphi(\delta^{is}f)} \;, \qquad \delta^{is}f(\vb*{x}) = f(\Lambda_{-s}\, \vb*{x}) \;, \label{da}
	\end{equation}
	where $\Lambda_s$ stands for a Lorentz boost: 
	\begin{align}
		\Lambda_s: 
		\left\{
		\begin {aligned}
		&  \;\; x'  =  \cosh(2\pi s) \;x - \sinh(2 \pi s) \; t, \\
		&  \;\;  t'  =  \cosh(2\pi s) \;t - \sinh(2 \pi s) \; x.          
	\end{aligned}
	\right. \label{bst}
\end{align}
The above equations enable us to evaluate the Araki-Uhlmann relative entropy. In fact, 
\begin{eqnarray} 
\langle \Psi |\; \Delta_{\psi|\Omega}^{is} \; |\Psi \rangle & = & \langle \Omega |\; {\cal A}_{-f} \Delta_{\Omega}^{is} {\cal A}_f \; |\Omega \rangle  
=  \langle \Omega |\; {\cal A}_{-f} \Delta_{\Omega}^{is} {\cal A}_f \Delta_{\Omega}^{-is}\; |\Omega \rangle = 
\langle \Omega |\; {\cal A}_{-f} {\cal A}_{(\delta^{is}f)} \; |\Omega \rangle  \nonumber \\
&=& e^{\frac{i}{2} \Delta_{PJ}(f, \delta^{is}f)} \; e^{-\frac{1}{2} ||f - \delta^{is}f||^2}  \;, \label{fev}
\end{eqnarray}
where we explored the invariance of the vacuum state $|\Omega\rangle$ under $\Delta_{\Omega}$. Therefore, taking the derivative with respect to $s$ and setting it to zero, one gets \cite{Ciolli:2019mjo,Casini:2019qst,Frob:2024ijk}
\begin{equation} 
S(\Psi |\Omega) = - \frac{1}{2} \Delta_{PJ}(f,f'_s|_{s=0}) \;, \label{fent}
\end{equation}
where $ f'_s|_{s=0}$ stands for 
\begin{equation} 
f'_s|_{s=0} = \frac{d}{ds} f(\Lambda_{-s}\,\vb*{x}) \Big|_{s=0} \;. \label{dss}
\end{equation}
Expression \eqref{fent} will be the starting point of our numerical analysis. 

%%%%%%%%%%%%%%%%%%%%%%%%%%%%%%%%%%%	
\section{Definition and localization of the test function}\label{sec.ftest}
%%%%%%%%%%%%%%%%%%%%%%%%%%%%%%%%%%%	

\begin{figure}[t!]
\centering
\begin{minipage}[b]{0.6\linewidth}
\includegraphics[width=\textwidth]{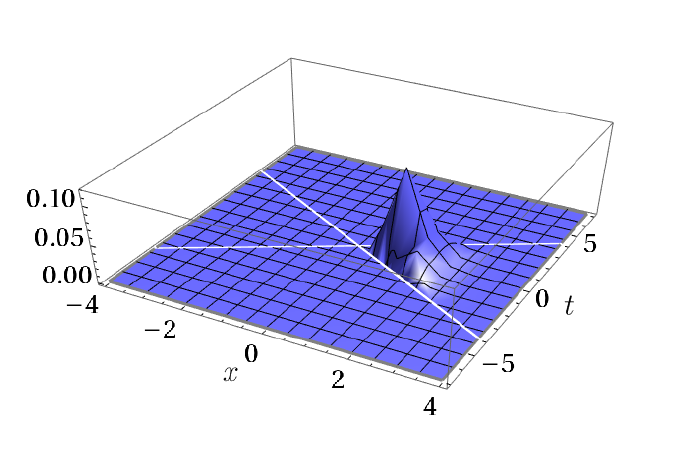}
\end{minipage} \hfill
\caption{Plot of the test function $f(t,x)$ for $\eta=1$ and $\alpha=4$.}
\label{ff}
\end{figure}

Let us now discuss the construction of the test function $f(t,x)$. Since it has to be a smooth function with compact support, we shall consider the following profile:
\begin{align}
f(t,x) = \eta
\left\{
\begin {aligned}
& e^{-\frac{1}{x^2-t^2}}\;e^{-\frac{1}{\alpha^2-(x^2-t^2)}}\;e^{-x^2}, \quad & \alpha \ge x \geq |t|,  \\
& 0,  \;\;\;\; {\rm elsewhere},                   
\end{aligned}
\right. \label{ftf}
\end{align}
where $\eta$ is a normalization factor and $ \alpha$ is a positive parameter. The behavior of $f(t,x)$ is depicted in Fig.~\eqref{ff}.

It is evident that $f(t,x)$ is properly localized in the right wedge, vanishing entirely in the left wedge, as required. Clearly, the factor $ e^{-\frac{1}{x^2-t^2}}$ plays a crucial role in restricting the support of $f(t,x)$ to the intended region. In addition, we have introduced the parameter $\alpha$, which appears in the term $e^{-\frac{1}{\alpha^2-(x^2-t^2)}}$. This parameter has the meaning of a slicing parameter: as $\alpha$ increases, the size of the support of $f(t,x)$ increases. Therefore, $\alpha$ provides a means to verify a key property of the Araki-Uhlmann relative entropy, given by \cite{Witten:2018zxz}
\begin{equation} 
S(\Psi |\Omega)(U) \ge S(\Psi |\Omega)(\tilde U) \;, \label{grw}
\end{equation}
where $(U, {\tilde U})$ stand for two spacetime regions, with ${\tilde U}$ contained within $U$. When expressed in terms of the parameter $\alpha$, Eq.~\eqref{grw} takes the form 
\begin{equation} \label{grwa}
S(\Psi |\Omega)(\alpha) \ge S(\Psi |\Omega)(\tilde \alpha) \;, \qquad \alpha \ge {\tilde \alpha} \;. 
\end{equation}
Finally, it should be noted that, according to Eq.~\eqref{da}, the test function $f(t, x)$ must depend on the boost parameter when acted upon by $\Lambda_{-s}$. This feature is accounted for by the term $e^{-x^2} $ which is not preserved under $\Lambda_{-s}$. Performing thus the boost transformation \eqref{bst}, the final form of the test function reads 
\begin{align}
f(t,x,s) = \eta
\left\{
\begin{aligned}
& e^{-\frac{1}{x^2-t^2}}\;e^{-\frac{1}{\alpha^2-(x^2-t^2)}}\;e^{- (\cosh(2\pi s) \;x + \sinh(2 \pi s) \; t )^2}, \quad & \alpha \ge x \geq |t|,  \\
& 0,  \;\;\;\; {\rm elsewhere}.                   
\end{aligned}
\right. \label{ftf1}
\end{align}
%which is essential for the computation of the relative entropy \eqref{fent} explicitly.

%%%%%%%%%%%%%%%%%%%%%%%%%%%%%%%%%%%	
\section{Numerical setup and results} \label{ns}
%%%%%%%%%%%%%%%%%%%%%%%%%%%%%%%%%%%	

We start this section by providing some details about the numerical setup we have designed. According to expression \eqref{fent}, the basic integral to be evaluated is $ \Delta_{PJ}(f,f'_s|_{s=0})$. Because deriving a closed analytical expression for the Fourier transform of the test function \eqref{ftf1} is challenging, the integral has been directly evaluated in configuration space (see Eq.~\eqref{mint}). The numerical integration has been performed with Mathematica by using the {\it Quasi-Montecarlo} method with precision set by {\it MaxPoints} $= 10^{8}$.

\begin{figure}[!t]
\begin{center}
\includegraphics[width=1.05\linewidth]{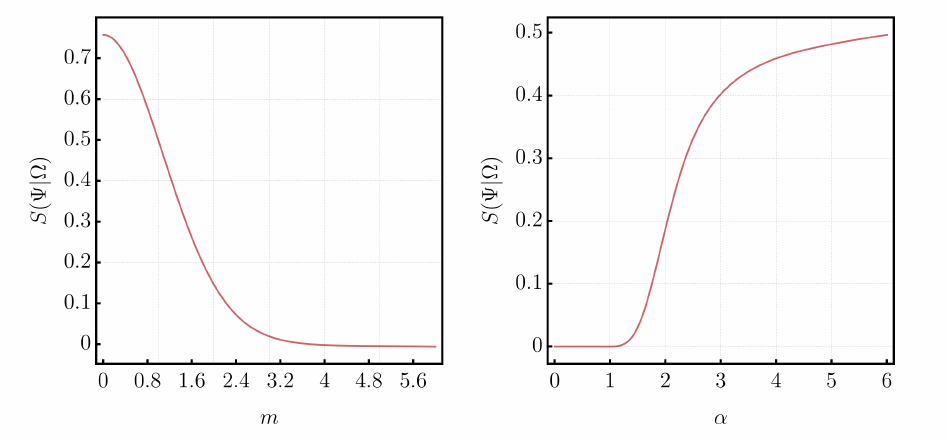}
\end{center}
\caption{\label{fig:PlotsEntropyWedge} We show the behavior of the Araki-Uhlmann relative entropy as a function of the mass parameter $m$ for $\alpha=6$ (left panel) and as a function of the slicing parameter $\alpha$ for $m=1$ (right panel) for the test function  $f(t,x)$, Eq.~\eqref{ftf}. For both plots we set $s=0$ and $\eta=4$.} 
\end{figure}

More specifically, the numerical computation of the Araki-Uhlmann relative entropy requires the evaluation of a spacetime integral of the form
\begin{equation} 
{\cal J}(m,\alpha) = \int d^2x \;d^2{x'} \;f(t,x) \Delta_PJ(t-t', x-x') f'_s(t',x')\Big|_{s=0} \;, \label{type} 
\end{equation} 
where, for definiteness, we adopt the test function specified in Eq.~\eqref{ftf1}. Expression \eqref{type} represents a four-dimensional integral, with the domain of integration determined by the support of the test function $f(t,x)$. To generate a plot analogous to the left panel of Fig.~\eqref{fig:PlotsEntropyWedge}, one begins by selecting a fixed value for the parameter  $\alpha$, for instance $\alpha = 6$. The integral is then numerically evaluated using the \textit{Quasi-Monte Carlo} method implemented in Mathematica, over a set of points uniformly distributed in the interval [0,5.6], as illustrated in Fig.~\eqref{fig:PlotsEntropyWedge}. The mass parameter $m$ is discretized with a step size of 0.01. This procedure yields an initial discrete (dotted) dataset. A continuous plot is subsequently obtained by interpolating the resulting data points. This summarizes the essential numerical strategy adopted in the computation. 

Since the scalar field is massive, the Araki-Uhlmann relative entropy shows a dependence on the mass parameter $m$ and on the slicing parameter $\alpha$. The first result addresses how the relative entropy varies with the parameter $m$, as shown in the left panel of Fig.~\eqref{fig:PlotsEntropyWedge}.
%\begin{figure}[t!]
%\centering
%	\begin{minipage}[b]{0.5\linewidth}
%		\includegraphics[width=\textwidth]{fig2.pdf}
%\end{minipage} \hfill
%\caption{Plot of the Araki-Uhlmann relative entropy as a function of the mass parameter $m$ for $(\eta=4, \alpha=6 )$.   }
%\label{me}
%\end{figure}

One can see that the entropy decreases as the mass parameter increases. This behavior is expected: as the mass grows, more degrees of freedom become heavy and effectively frozen, leading to a reduction in entropy.

The second result concerns the dependence on the slicing parameter $\alpha$, as shown in the right panel of Fig.~\eqref{fig:PlotsEntropyWedge}. The entropy increases with the growth of the slicing parameter $\alpha$, in full agreement with Eq.~\eqref{grwa}. Initially, the increase is slow, but then it accelerates.

\begin{figure}[t!]
\centering
\begin{minipage}[b]{0.5\linewidth}
\includegraphics[width=\textwidth]{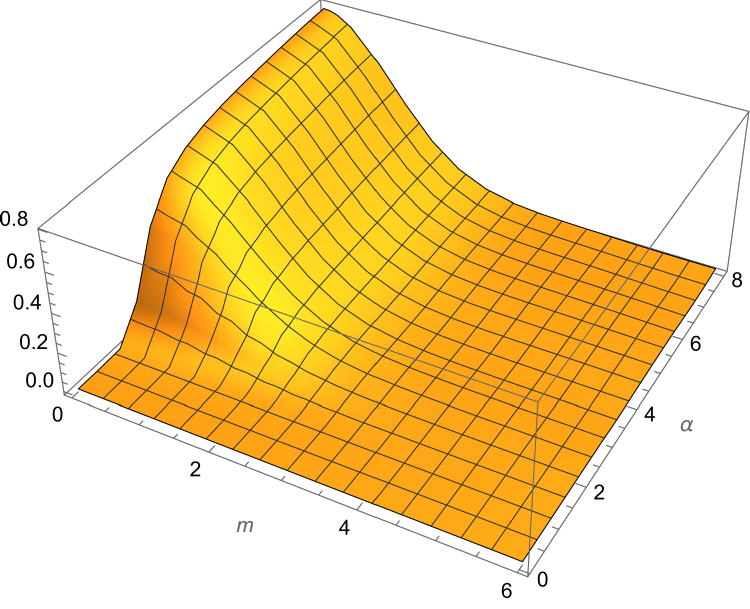}
\end{minipage} \hfill
\caption{Three-dimensional plot of the Araki-Uhlmann entropy as a function of the slicing parameter $\alpha$ and of the mass $m$, for $\eta=4 $.   }
\label{fig4}
\end{figure}

In Fig.~\eqref{fig4}, a three-dimensional plot of the Araki-Uhlmann relative entropy as a function of the parameters $(\alpha, m)$ is shown. Notably, the positivity of the entropy can be observed.

Finally, we note that the parameter $\eta$ does not significantly influence the behavior of the Araki-Uhlmann relative entropy. It only determines the overall magnitude of the relative entropy, without affecting its dependence on $m$ and $\alpha$. 
%\begin{figure}[t!]
%\centering
%\begin{minipage}[b]{0.5\linewidth}
%\includegraphics[width=\textwidth]{fig3.pdf}
%\end{minipage} \hfill
%\caption{Plot of the Araki-Uhlmann relative entropy as a function of the slicing parameter $\alpha$ for $(\eta=4, m=1 )$.   }
%\label{ae}
%\end{figure}

\subsection{Analysis with different test functions}\label{testchange}

One could question whether altering the test function would lead to changes in the behavior of the relative entropy. To address this, we can explore the use of different test functions.  
%such as the one shown in Fig.~\eqref{disk1}.
%\begin{figure}[!h]
%\centering
%\begin{minipage}[b]{0.5\linewidth}
%	\includegraphics[width=\textwidth]{PlotfTest2.pdf}
%\end{minipage} \hfill
%\caption{Plot of the test function $f_{d} $ supported in the disk $(x-b)^2+t^2 \le a^2$.  The plot refers to $a=1.5$, $b=8$. }
%\label{disk1}
%\end{figure}
\subsubsection{Second alternative: disk-supported test function}
For this purpose, we choose a second test function $f_2(t,x)$ which is supported in the disk of radius $a$ and centered at the point $(b,0)$,
\begin{equation} 
(x-b)^2 +t^2 \le a^2 \;, \label{ds}
\end{equation}
where the parameters $(a,b)$ are chosen so that the disk lies within the right Rindler wedge.% cf. Fig.~\eqref{disk2}.
%\begin{figure}[h]
%\centering
%	\begin{minipage}[b]{0.5\linewidth}
%		\includegraphics[width=\textwidth]{DiskRegion1.pdf}
%	\end{minipage} \hfill
%\caption{The support of the test function $f_d(t,x)$ is a disk centered at the point $b$ with a radius $a$. For this plot, we have set $a=3.5$ and $b=4$.}
%	\label{disk2}
%	\end{figure}

\begin{figure}[t]
\begin{center}
\includegraphics[width=1.05\linewidth]{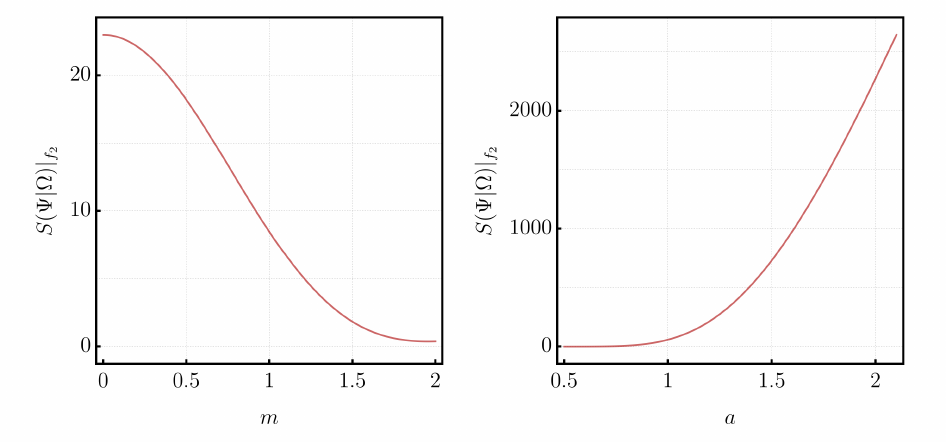}
\end{center}
\caption{\label{fig:PlotsEntropyDisk} We show the behavior of the Araki-Uhlmann relative entropy as a function of the mass parameter $m$ for $a=2$ and $b=200$ (left panel) and as a function of the radius $a$ for $m=0.1$ and $b=200$ (right panel) for the test function  $f_2(t,x)$, Eq.~\eqref{fd}. For both plots we set $s$ to zero.}
\end{figure}

\noindent For the profile of the test function $f_2$, we choose 
\begin{align}
f_2(t,x) = 
\left\{
\begin {aligned}
& e^{-\frac{1}{x^2-t^2}}\;e^{-\frac{1}{a^2-(x-b)^2-t^2}}\;e^{-(x-b)^2}, \quad & (x-b)^2 + t^2 \le a^2,  \\
& 0,  \;\;\;\; {\rm elsewhere}.                   
\end{aligned}
\right. \label{fd}
\end{align}
The behavior of the relative entropy as a function of the mass $m$ and the radius $a$ of the disk is shown in Fig.~\eqref{fig:PlotsEntropyDisk}. 
%It can be observed that no qualitative changes occur.

The behavior of the Araki-Uhlmann relative entropy for a test function supported in a disk follows a pattern similar to the case where the test function is localized in the right Rindler wedge. The relative entropy decreases as $m$ increases, which aligns with expectations since larger masses suppress quantum fluctuations, reducing entanglement contributions. In contrast, the relative entropy increases with the radius $a$ of the disk, as a larger spatial region captures more quantum correlations, leading to greater entropy. This trend mirrors the behavior observed in the Rindler wedge case, where the entropy grows as the support of the test function expands. 

We emphasize that the parameter $b$ (the center of the disk) cannot be chosen smaller than the radius $a$. Such a choice would position the test function, particularly its derivative, too close to the Rindler horizon, where the Pauli-Jordan function exhibits significant oscillations, resulting in excessive instability.

\subsubsection{Third alternative: smooth diamond-supported test function}
\begin{figure}[t]
\begin{center}
\includegraphics[width=1.05\linewidth]{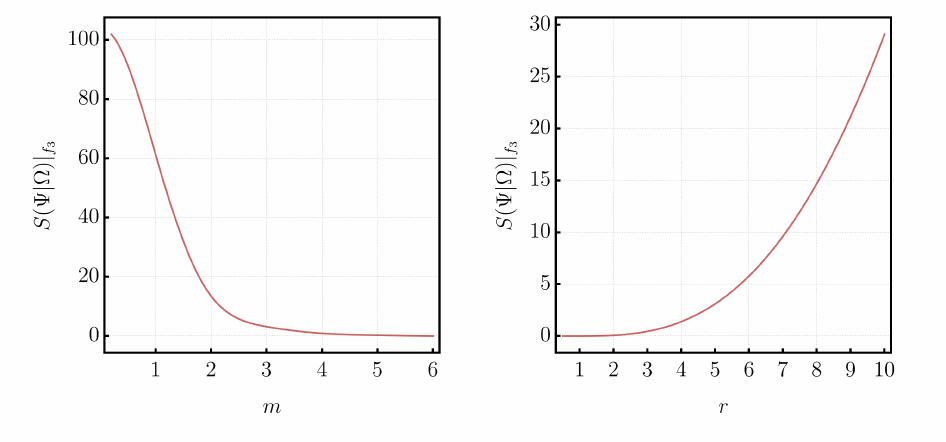}
\end{center}
\caption{\label{fig:PlotsEntropyf3} Plot of the Araki-Uhlmann relative entropy as a function of the mass parameter $m$ for $r=2$, $d=5$ and $\eta=1$ (left panel) and as a function of the parameter $r$ for $d=50$, $m=0.1$ and $\eta=10^{-2}$ (right panel) for the test function  $f_3(t,x)$, Eq.~\eqref{f3}. For both plots we set $s$ to zero.}
\end{figure}

For the third type of test function, we introduce a smooth variant of the test function supported within the causal diamond
$|d+r-x|+|t|\leq r$, where $d$ denotes the center of the diamond, $r$ is a parameter that controls its size and $x \in [d,\, d+2r]$. To achieve smoothness, we define the auxiliary variable
\begin{equation}
\lambda(\epsilon)=\sqrt{(d+r-x)^2+\epsilon^2}+\sqrt{t^2+\epsilon^2},
\end{equation}
where the smoothness is governed by the parameter $\epsilon$. From a numerical perspective, we choose $\epsilon=10^{-3}$. The shape of the third test function is then given by
\begin{align}
f_3(t,x)=\eta
\left\{
\begin{aligned}
& {\rm exp}\bigg[{-\frac{1}{(\sqrt{r^2+\epsilon^2}+\epsilon)^2-\lambda(\epsilon)^2}}\bigg] \;, \;\;\;\;\;
& \lambda(\epsilon)\leq \sqrt{r^2+\epsilon^2}+\epsilon,\;\\
&0,\;\;\;\; {\rm elsewhere}.
\end{aligned}
\right. \label{f3}
\end{align} 
Here, $\eta$ serves as a normalization factor. Fig.~\eqref{fig:PlotsEntropyf3} shows the behavior of the Araki-Uhlmann relative entropy for the test function $f_3(t,x)$ with respect to its parameters. As shown in both panels, this test function exhibits similar qualitative behavior to the previous cases: decreasing with mass and increasing with the size parameter $r$, with the latter showing an initially gradual rise that becomes more pronounced at larger values.

\subsubsection{Fourth alternative: vertical strip-supported test function}
\begin{figure}[H]
\centering
\begin{minipage}[b]{0.5\linewidth}
\includegraphics[width=\textwidth]{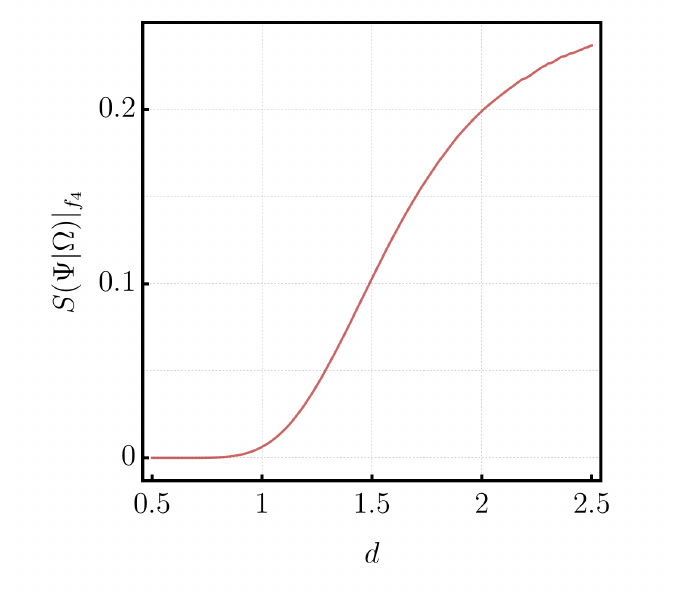}
\end{minipage} \hfill
\caption{Behavior of the relative entropy with respect to the slicing parameter $d$, for $m=0.1$,  $\beta=0.5$ for the test function $f_4(t,x)$, Eq.~\eqref{fbeta}.}
\label{betad}
\end{figure}

Finally, we introduce a fourth type of test function, which is supported in a vertical strip within the right Rindler wedge. It is defined as
\begin{align}
f_{4}(t,x) = 
\left\{
\begin {aligned}
& e^{-\frac{1}{x^2-t^2}}\;e^{-\frac{1}{((x^2-t^2)-\beta^2)((\beta+d)^2 -(x^2-t^2))}}\;e^{-(x-b)^2}\;e^{-t^2} \;, \;\; & (\beta+d)^2 \ge x^2-t^2\ge \beta^2 \;,  \;\;\;x \ge |t| \;,  \\
& 0,  \;\;\;\; {\rm elsewhere}.                   
\end{aligned}
\right. \label{fbeta}
\end{align}
Here, $\beta$ serves as the localization parameter, while $d$ controls the slicing. Increasing $d$ expands the region where $f_4(t,x)$ remains nonzero. The dependence of the relative entropy on the mass parameter follows the same trend observed previously. Regarding its behavior with respect to the slicing parameter, it is illustrated in Fig.~\eqref{betad}. %Overall, no qualitative changes arise from modifying the test functions.	

Overall, the relative entropy follows a universal pattern: it decreases with mass due to the suppression of fluctuations and increases with region size as more correlations are included. Altogether, the analysis of all test functions considered reinforce the general trends of the relative entropy’s dependence on mass and spatial extent, showing that different test function profiles do not alter these fundamental properties.

\section{Conclusion}\label{conc}

This work presented a detailed numerical investigation of the Araki-Uhlmann relative entropy in quantum field theory, focusing on the case of a coherent state relative to the vacuum in a free massive scalar field in $1+1$-dimensional Minkowski spacetime. The analysis employed a framework grounded in Tomita–Takesaki modular theory and systematically explored various classes of test functions with compact support, including configurations localized in the right Rindler wedge, a disk, a smooth diamond, and a vertical strip.

Across all test function profiles, the relative entropy was found to exhibit two robust behaviors: a monotonic decrease with the mass parameter $m$, and a monotonic increase with the size of the spacetime region defined by the test function's support (as quantified by parameters such as the slicing parameter $\alpha$, the disk radius $a$, or their analogues in other geometries). These trends are consistent with physical intuition—larger mass suppresses quantum fluctuations, reducing entanglement, while larger regions capture more quantum correlations.

Importantly, although these behaviors are theoretically anticipated, to our knowledge no formal proof exists in the literature. Thus, this work provides an explicit numerical exploration and confirmation of these expectations within the framework of free field theory. The observed universality—i.e., the independence of qualitative features from the specific choice of test function—underscores the structural robustness of the Araki-Uhlmann relative entropy as a measure of quantum distinguishability and entanglement.

%This work presented a detailed numerical investigation of the Araki-Uhlmann relative entropy in QFT, specifically focusing on the relative entropy between a coherent state and the vacuum state for a free massive scalar field in $1+1$ Minkowski spacetime. The study employed a numerical setup based on Tomita-Takesaki modular theory and explicit test functions profiles, and has revealed a universal behavior governing the Araki-Uhlmann relative entropy dependence on key parameters.

%Across all test function profiles studied — including those supported in the right Rindler wedge, a disk, a smooth diamond, and a vertical strip — the relative entropy consistently decreases as the mass parameter $m$ increases. This trend aligns with physical intuition, as larger masses suppress quantum fluctuations, leading to a reduction in entanglement contributions. Conversely, the relative entropy exhibits a monotonic increase as the spatial extent of the test function's support grows, whether characterized by the slicing parameter $\alpha$, the radius $a$, or analogous measures in other cases. This behavior confirms the expected result that larger regions capture more entanglement, reinforcing the positivity of the relative entropy and the interpretation of it as a robust measure of quantum distinguishability and entanglement in QFT.

%Importantly, while the specific numerical values depend on the precise choice of the test function, the overarching trends remain unchanged, demonstrating the universality of these entropy properties.

Looking ahead, this framework opens exciting avenues for future research. One particularly intriguing direction involves using test functions localized in diamond-shaped regions, as employed in recent investigations of the Bell-CHSH and Mermin inequalities \cite{Guimaraes:2024xtj,Guimaraes:2025xij}, to enable investigations into the entanglement properties of multi-coherent-state configurations, as discussed in \cite{Casini:2019qst}. In particular, an explicit exploration involves constructing a chain of coherent states distributed across diamond-shaped regions within the right and left wedges. This investigation appears feasible with an appropriate adaptation of the current numerical setup and is currently underway.

The results of this work not only deepen our understanding of the Araki-Uhlmann relative entropy in the context of QFT theory, but also pave the way for future explorations into the rich interplay between quantum information and relativistic systems. This numerical approach serves as a foundational tool for further investigations, with potential applications in diverse areas, including holography, thermodynamics, and the study of entanglement structure in curved spacetime geometries and in interacting field theories.

Finally, we briefly address the highly nontrivial problem of evaluating the Araki-Uhlmann relative entropy in the context of interacting QFTs. In principle, for theories defined on $1+1-$dimensional Minkowski spacetime, one may exploit the powerful framework of bosonization, particularly in models such as the massive Thirring model, which is dual to the sine-Gordon theory~\cite{coleman1975quantum,GomezNicola:1998pf}. The availability of exact correlation functions in such settings offers a promising avenue to analyze the dependence of the relative entropy on the interaction coupling constant. Any significant advancements in this direction will be communicated in due course.

\section*{Acknowledgments}
The authors would like to thank the Brazilian agencies Conselho Nacional de Desenvolvimento Científico e Tecnológico (CNPq), Coordenação de
Aperfeiçoamento de Pessoal de Nível Superior - Brasil (CAPES) and Fundação Carlos Chagas Filho de Amparo à Pesquisa do Estado do Rio de Janeiro (FAPERJ) for financial support. In particular, A. F.~Vieira is supported by a postdoctoral grant from FAPERJ, grant No. E-
26/200.135/2025. S. P.~Sorella, I.~Roditi, and M. S.~Guimaraes are CNPq researchers under contracts 301030/2019-7, 311876/2021-8, and 309793/2023-8, respectively. 

%\end{acknowledgments}

\appendix

\section{The massive real scalar field in $1+1$ Minkowski spacetime}\label{appA}
In this appendix, we summarize some of the key aspects of the canonical quantization of a massive real scalar field in $1+1$-dimensional Minkowski spacetime.
\subsection{ Field expansion and commutation relations}

The massive scalar field $\varphi(\vb*{x})$ can be expressed in terms of plane waves as:
\begin{equation} \label{qf}
\varphi(\vb*{x}) = \int \! \frac{d k}{2 \pi} \frac{1}{2 \omega_k} \left( e^{-ik_\mu x^\mu} a_k + e^{ik_\mu x^\mu} a^{\dagger}_k \right), 
\end{equation} 
where $\omega_k  = k^0 = \sqrt{k^2 + m^2}$ is the relativistic energy dispersion relation. The field operators satisfy the canonical commutation relations:
\begin{align}
[a_k, a^{\dagger}_q] &= 2\pi \, 2\omega_k \, \delta(k - q), \\ \nonumber 
[a_k, a_q] &= [a^{\dagger}_k, a^{\dagger}_q] = 0. 
\end{align}
Since quantum fields are operator-valued distributions \cite{Haag:1992hx}, they must be smeared with test functions to produce well-defined operators in Hilbert space. This is done by defining the smeared field operator as:
\begin{align} 
\varphi(h) = \int \! d^2\vb*{x} \; \varphi(\vb*{x}) h(\vb*{x}) \;, \label{smmd}
\end{align}
where $h(\vb*{x})$ is a real smooth test function with compact support. 

\subsection{Inner product and two-point functions}

With the smeared fields, the Lorentz-invariant inner product between two test functions $f(\vb*{x})$ and $g(\vb*{x})$ in the vacuum state is introduced by means of the two-point smeared Wightman function
\begin{align} \label{InnerProduct}
\langle f \vert g \rangle &= \langle 0 \vert \varphi(f) \varphi(g) \vert 0 \rangle =  \frac{i}{2} \Delta_{PJ}(f,g) +  H(f,g) \;, 
\end{align}
where $ \Delta_{PJ}(f,g)$ and $H(f,g)$ are the smeared versions of the Pauli-Jordan and Hadamard distributions, respectively. These are defined as:
\begin{align}
\Delta_{PJ}(f,g) &=  \int \! d^2\vb*{x} d^2\vb*{y} f(\vb*{x}) \Delta_{PJ}(\vb*{x}-\vb*{y}) g(\vb*{y}) \;,  \nonumber \\
H(f,g) &=  \int \! d^2\vb*{x} d^2\vb*{y} f(\vb*{x}) H(\vb*{x}-\vb*{y}) g(\vb*{y})\;. \label{mint}
\end{align}
The Pauli-Jordan function $\Delta_{PJ}(\vb*{x}-\vb*{y})$ and the Hadamard function $H(\vb*{x}-\vb*{y})$ take the explicit forms:
\begin{eqnarray} 
\Delta_{PJ}(t,x) & =&  -\frac{1}{2}\;{\rm sign}(t) \; \theta \left( \lambda(t,x) \right) \;J_0 \left(m\sqrt{\lambda(t,x)}\right) \;, \nonumber \\
H(t,x) & = & -\frac{1}{2}\; \theta \left(\lambda(t,x) \right )\; Y_0 \left(m\sqrt{\lambda(t,x)}\right)+ \frac{1}{\pi}\;  \theta \left(-\lambda(t,x) \right)\; K_0\left(m\sqrt{-\lambda(t,x)}\right) \;, \label{PJH}
\end{eqnarray}
where 
\begin{equation} 
\lambda(t,x) = t^2-x^2 \;, \label{ltx}
\end{equation}
and $(J_0,Y_0,K_0)$ are Bessel functions, while $m$ is the mass parameter.

\subsection{Physical interpretation and causality}

Both the Hadamard and Pauli-Jordan distributions are Lorentz-invariant. In particular, the Pauli-Jordan distribution, $\Delta_{PJ}(\vb*{x})$, plays a fundamental role in encoding relativistic causality, as it vanishes outside the light cone. Additionally, $\Delta_{PJ}(\vb*{x})$ and the Hadamard distribution, $H(\vb*{x})$, exhibit distinct symmetry properties: $\Delta_{PJ}(\vb*{x})$ is antisymmetric under the transformation $\vb*{x} \to -\vb*{x}$, while $H(\vb*{x})$ remains symmetric. When expressed in terms of smeared fields, the commutator of the field operators is given by  
\[
\left[\varphi(f), \varphi(g)\right] = i \Delta_{PJ}(f, g).
\]  
This ensures that causality is preserved, as two field operators commute whenever the supports of $f(\vb*{x})$ and $g(\vb*{x})$ are spacelike separated:  
\[
\left[\phi(f), \phi(g)\right] = 0.
\]

\bibliography{refs}
%
%%%%%%%%%%%%%%%%%%%%%%%%%%%%%%%%%%%%%%%%%%%%%%%%%%%%%%%%%%%%%%%%%%%%%%%%%%%%%%%%%%%%%%%%%%%%%%%%%%%% 

\end{document}